%% file: main.tex
\documentclass[conference, 10pt]{IEEEtran}
\IEEEoverridecommandlockouts
\usepackage{cite}
\usepackage{amsmath,amssymb,amsfonts}
\usepackage{graphicx}
\usepackage{textcomp}
\usepackage[table,xcdraw]{xcolor}
\def\BibTeX{{\rm B\kern-.05em{\sc i\kern-.025em b}\kern-.08em
    T\kern-.1667em\lower.7ex\hbox{E}\kern-.125emX}}

\usepackage{algorithm, algpseudocode}
\usepackage{array}

\usepackage[caption=false,font=normalsize,labelfont=sf,textfont=sf]{subfig}
\usepackage{stfloats}
\usepackage{float}

\usepackage{url}
\usepackage{booktabs}
\usepackage{multirow}

\begin{document}

\title{Optimizing Vehicular Users Association in Urban Mobile Networks}

\author{Geymerson S. Ramos$^{1,2}$, Razvan Stanica$^2$, Rian G. S. Pinheiro$^1$, Andre L. L. Aquino$^1$\\
geymerson@laccan.ufal.br, razvan.stanica@inria.fr, rian@ic.ufal.br, alla@laccan.ufal.br\\
$^1$Computing Institute, Universidade Federal de Alagoas\\
Av. Lourival Melo Mota, S/N, Tabuleiro do Martins, 57072-970, Maceió, Alagoas, Brazil\\
$^2$Univ Lyon, Inria, INSA Lyon, CITI - 56 Bd Niels Bohr, 69100, Villeurbanne, France
}

\maketitle

\begin{abstract}

This study aims to optimize vehicular user association to base stations in a mobile network. We propose an efficient heuristic solution that considers the base station average handover frequency, the channel quality indicator, and bandwidth capacity. We evaluate this solution using real-world base station locations from São Paulo, Brazil, and the SUMO mobility simulator. We compare our approach against a state of the art solution which uses route prediction, maintaining or surpassing the provided quality of service with the same number of handover operations. Additionally, the proposed solution reduces the execution time by more than 80\% compared to an exact method, while achieving optimal solutions.
\end{abstract}


\section{Introduction} 
\label{sec:introduction}

Mobile networks are a highly dynamic, dense, and connected ecosystem. Users seamlessly transition between environments and communication technologies through handover processes. Handovers occur when user equipment (UE) changes from one access point, base station, or evolved Node B (eNB) to another. While moving, the system must sustain the users' connections without them perceiving communication discontinuity or a drop in transfer rates. In this sense, Reference Signal Received Quality (RSRQ) is a crucial indicator of communication quality between users and base stations, commonly used in the handover decision process.

However, handovers come with a price in terms of user Quality of Service (QoS). Every time a handover takes place, the connection between the UE and the radio access network (RAN) needs to be reestablished. Moreover, the packets already transmitted at the old eNB need to be re-routed towards the new one, which also implies a certain delay. If the overall handover execution duration is too significant, the communication might be disrupted and the user QoS drops. Therefore, when the RAN periodically takes the user association decision, it has to consider both the RSRQ gain brought by a handover and the impact on QoS this implies.

This problem is particularly stringent for vehicular users, who present a relatively high level of mobility. Moreover, if this mobility takes place in an urban environment, where the eNB deployment is dense, the user association decision becomes even more complex. Minimizing the number of handovers in the system~\cite{ramos20195Gsdn} and the communication cost between base stations and data centers~\cite{tarik2015} represent major objectives for mobile operators. These optimization problems are usually tackled through integer linear programming (ILP), which allows obtaining optimal solutions. On the downside, ILP generally comes with high computational costs, and it can rarely be used for online decisions in networking problems.



In this paper, we propose an efficient meta-heuristic solution for the vehicular user association problem. A heuristic approach is a better alternative because it computes network solutions much faster than an ILP model, making it suitable for online deployment in the context of 5G and software-defined networks. We utilize the user RSRQ as an objective metric. However, in order not to trigger a QoS decrease by producing too many handovers, we also include the UE handover frequency in the optimization problem. Our heuristic is based on an iterated local search algorithm, and we evaluate it against the ILP formulation proposed in~\cite{ramos20195Gsdn}, which produces an optimal solution.

The proposed heuristic approach is beneficial in the case of the user association problem, which is continuously solved at the RAN level, with computation requirements in the order of a few milliseconds. A suboptimal user association decision can be tolerated from time to time, if the QoS degradation with respect to the optimal strategy is not too significant. Indeed, in our results, we observe that the iterated local search algorithm successfully discovers optimal (or close to optimal) solutions with reduced execution time compared to the ILP model. Moreover, we compare our solution against the multi-route prediction allocation model~\cite{ahmadi:2020}, and we notice a better QoS, by improving the RSRQ indicator, while performing the same number of handovers. To summarize, our main contributions are as follows: \emph{i)} the design of an efficient heuristic to allocate vehicular users to base stations in mobile networks; and \emph{ii)} the presentation of a mobility simulation methodology that considers the real-world base station locations in an urban area, provided by a local mobile carrier. We provide the source code and data for the discussed models in \cite{sourcecode2023}.


The remainder of this paper is structured as follows: 
Section~\ref{sec:related_work} presents the related work. 
Section~\ref{sec:alloc_model} describes the RSRQ user association model and the heuristic solution. 
Section~\ref{sec:results} and Section~\ref{sec:conclusion} present the results and final considerations, respectively.

\section{Related Work}\label{sec:related_work}


The abundance of network services poses a challenge in creating a general optimization model for tasks like handover decision making, user association, and bandwidth management~\cite{xu2021survey}. For example, Lee et al.~\cite{jiseong:2017} propose a 5G handover scheme that considers users mobility information, including speed, movement direction, the angle between UE and the base station antenna, and communication distance. Their study utilizes a software-defined network (SDN) controller to determine the suitable base station for user association, resulting in reduced message exchanges and improved load management within cells. In contrast, our current model does not require speed or direction data to decide on user association.

Bakht et al.~\cite{bakht2019powerAllocation} propose a power allocation and user association scheme for heterogeneous networks. They aim to ensure fairness among users in network services, by solving this joint problem. They design an algorithm based on dual decomposition~\cite{caroe1999dualDecomp} to find optimal solutions, but this approach is resource-intensive and time-consuming. Moreover, this work needs a refined mobility simulation model to evaluate the impact of handover decisions. 

Taleb et al.~\cite{tarik2015} propose a multi-objective model for minimizing the number of handovers and communication costs in wireless networks. They consider the Evolved Packet System (EPS) architecture in an SDN context, using virtual network infrastructure components replacing the Serving Gateway (S-GW) and Packet Gateay (P-GW). Virtual function allocation is based on network usage and user behavior, with P-GW functions placed in data centers closer to eNBs to reduce communication costs. To minimize the number of handovers, S-GW functions are allocated to expand the coverage areas. However, this approach requires a complex multi-objective model and does not address user association to eNBs.

Ahmadi et al.~\cite{ahmadi:2020} propose a solution using RSRQ and a prediction factor based on the potential routes of the users to their destination. One drawback is the need for destination knowledge and computing all possible routes for each user, which is a $\mathcal{NP}$-hard problem~\cite{Chatterjee2014PossiblePaths}. In our prior work~\cite{ramos20195Gsdn}, we presented a solution considering variables like UE-eNB distance, average handover frequency, and bandwidth requirements. Our contribution improves~\cite{ahmadi:2020} by incorporating RSRQ as a target metric and eliminating route prediction. In this work, compared to~\cite{ramos20195Gsdn}, we provide a heuristic method for user association, significantly reducing the computation cost. 


To summarize, Table~\ref{tbl:qualitative_analysis} compares our contribution and the mentioned works. This table focuses on five comparison criteria, considered or not in the corresponding study: QoS, bandwidth management (BM), mobility simulation (MS), user association model (UAM), and heuristic solution (HS).


\begin{center}
    \input{tables/qualitative_table}
\end{center}

\section{Proposed Model}
\label{sec:alloc_model}

From a general point of view, we need to assign users to base stations according to a number of constraints. This user association problem is a specific instance of the Generalized Assignment Problem (GAP)~\cite{cattrysse1992survey}. Each user must connect to one base station, and each base station has a maximum bandwidth capacity. The user association problem considers the available bandwidth, the user RSRQ, and the average handover frequency of eNBs. The average handover frequency parameter associates users to eNBs less likely to perform another transfer in a short interval. The bandwidth constraints help the model execution by balancing the network load and achieving base station capacity management.

\subsection{Exact Approach}

Consider the bipartite graph $G = (V, E)$ with initial eNB and UE configuration. Vertices $V = N \cup U$ consist of eNBs $N = \{eNB_1, eNB_2, \dots, eNB_n\}$ and UEs $U = \{UE_1, UE_2, \dots, UE_k\}$. Network connections between users and base stations are represented by edges $E = \{e_1, e_2, \dots, e_{(nk)}\}$, and each edge is associated with a communication cost $C = \{c_1, c_2, \dots, c_{(nk)}\}$. The proposed model aims to optimize the network configuration $S$ from the initial configuration $G$ by minimizing the global cost and ensuring the connection of each UE to the best eNB. To achieve this, we use an exact approach that establishes connections between users and base stations based on the best RSRQ value and the lowest handover frequency. The RSRQ metric provides noise and interference levels information, indicating received signal quality. This approach can be formally defined as follows:
{\allowdisplaybreaks
\begin{align}
    \min~z(b_{ki}, \overline{h_i}, \Theta_{ki}) = &\sum_{k \in U}{\sum_{i \in N}{b_{ki}\, (\overline{h_i} + \Theta_{ki})}} &\label{equ:users}\\
\text{subject to} \quad & \sum_{i \in N} {b_{ki} = 1} &\forall ~k \in U\label{equ:rest6}\\
    & \sum_{k \in U} {l_{ki}\, b_{ki} \leq L_i} &\forall ~i \in N \label{equ:rest7} \\
    & b_{ki} \in \mathbb B &\forall ~i \in N \label{equ:rest8}
\end{align}}
where $b_{ki}$ indicates if UE$_k$ is associated to eNB$_i$ or not, and $\overline{h_i}$ is the average handover frequency of the same eNB, defined as: 
\begin{align}
    & \overline{h_i} = ~\sum_{j \in N\setminus\{i\}} {\frac{h_{ij}}{(|N| - 1)}}        &\forall ~i \in N.
\label{equ:havg}
\end{align}
The average handover frequency between eNB$_i$ and eNB$_j$ is denoted as $h_{ij}$, and $|N|$ is the total of eNBs. Constraint~\eqref{equ:rest6} ensures that each user connects with only one eNB. Constraint~\eqref{equ:rest7} limits the maximum bandwidth capacity $L_i$ of eNB$_i$ to the sum of the bandwidth minimum requirements $l_{ki}$ for each user $k \in U$. Constraint~\eqref{equ:rest8} defines the binary domain of $b_{ki}$. Finally, the RSRQ between UE$_k$ and eNB$_i$, is defined as:
\begin{align}
    \Theta_{ki} =- \left[ \frac{d_{ki}}{R_i}(|\Theta_{min}|-|\Theta_{max}|)+|\Theta_{max}|\right],
\label{equ:rsrq}
\end{align}
where $d_{ki}$ represents the user-to-base station distance, $R_i$ is the maximum transmission radius of eNB$_i$, and $\Theta$ denotes the RSRQ value in dB. Signal degradation occurs as distance increases. RSRQ quality classification has four levels: Excellent ($\geq -5$ dB), Good ($-5, -9$] dB, Fair ($-9, -12$] dB, and Poor ($<-12$ dB). Poor reception typically indicates users located further from the eNB or at cell edges, while excellent or good reception suggests proximity to the base station. From Eq.~\ref{equ:rsrq} and RSRQ indicators, $\Theta_{ki} = -12$ dB when $d_{ki} = R_i$ (poor signal quality). Conversely, when $d_{ki} = 0$, $\Theta_{ki} = -5$ dB.


\subsection{Heuristic Approach}\label{ssec:userAlloc2}

Our heuristic method utilizes the iterated local search (ILS) metaheuristic~\cite{Lourencco2019Iterated} with variable neighborhood descent (VND)~\cite{Hansen_2018} in the local search phase. Additionally, we define two neighborhood structures for swap and insertion operations, ensuring compliance with constraints. The \texttt{swap\_ue($S$)} neighborhood swaps a user UE$_u$ associated to eNB$_i$ with a user UE$_v$ associated to eNB$_j$, halting at the first swap that improves the current solution. Its worst-case time complexity for $k$ UEs is $\mathcal{O}(k^2)$. The \texttt{insert\_ue($S$)} neighborhood removes a user UE$_u$ associated to eNB$_i$ and inserts it into the list of eNB$_j$, stopping at the first insertion that enhances the current solution. Its worst-case time complexity for $k$ UEs and $n$ eNBs is $\mathcal{O}(nk)$. The evaluations of cost improvement for these neighborhood structures, with $\mathcal{O}(1)$ complexity, are efficient as we directly access values in our model data structures.

Algorithm~\ref{alg:ils} presents our ILS metaheuristic. We start with an initial solution $S_0$ (Algorithm~\ref{alg:initSol}), likely a local minimum network configuration. The local search improves solutions using insert and swap operations. Line 2 updates $S_0$ to a new variable $S$ (Algorithm~\ref{alg:vndLocalSearch}). The ILS loop searches for improvements until a time stop condition $t \leq T$. The perturbation function at Line 6 introduces random swaps or insertions to escape local minimums. Line 7 uses the perturbed solution $S_{p}$ as input for the local search. If \texttt{Cost}($S_{p^*}$) $\leq$ \texttt{Cost}($S$) (Line 9), $S_{p^*}$ becomes the new best solution $S$. If $iter \geq \alpha$ with no improvement, a new initial solution $S$ is generated (Line 12).

\begin{algorithm}[!htb]
    \caption{Iterated Local Search - ILS}
    \label{alg:ils}
    \begin{footnotesize}
    \begin{algorithmic}[1]
        \Require The user connected network $G$
        \Ensure A new network configuration solution $S$
          \State{$S_0 \gets$ \texttt{GenInitialSolution}$(G)$}  \Comment{Algorithm~\ref{alg:initSol}}
          \State{$S \gets$ \texttt{VNDLocalSearch}$(S_0)$}  \Comment{Algorithm~\ref{alg:vndLocalSearch}}
          \State{$iter \gets$ $0$}
          \State{$t \gets$ $0$}
            \While {\textit{($t \leq T$)}}          
         	    \State{$S_p \gets$ \texttt{Perturbation}$(S)$}
          	    \State{$S_{p*} \gets$ \texttt{VNDLocalSearch}$(S_p)$}
          	    \State{$iter \gets$ $iter+1$}
          	    \If{$\text{\texttt{Cost}}(S_{p^*}) \leq \text{\texttt{Cost}}(S)$}
         	        \State{$S \gets S_{p^*}$}
         	        \State{$iter \gets$ $0$}
         	    \ElsIf{$iter \geq \alpha$}
         	        \State{$S \gets$ \texttt{GenInitialSolution}$(G)$}
         	        \State{$iter \gets$ $0$}
         	   \EndIf
         	\EndWhile
    \end{algorithmic}
    \end{footnotesize}
\end{algorithm}

Algorithm~\ref{alg:initSol} randomly assigns UEs to eNBs based on minimum cost. It randomizes the indexes of UEs and eNBs to change the allocation order (line 1). Randomization utilizes the Mersenne Twister generator~\cite{matsumoto1998mersennetwister} with uniform distributions. All network users are selected (Line 2) and then connected to their optimal eNB (Lines 4-8).

\begin{algorithm}[!htb]
    \caption{Initial Solution Generation}
    \label{alg:initSol}
    \begin{footnotesize}
    \begin{algorithmic}[1]
        \Require The user connected network $G$
        \Ensure An improved network configuration $S_0$
            \State{$S_0 \gets$ \texttt{UEandEnbIndexRandomization$(G)$}}
            \State{$UE \gets$ \texttt{NetworkUsers($S_0$)}}
            \State{$u \gets 0$}
            \Repeat
                \State{$eNB \gets$ \texttt{BestBaseStation}($UE_{u}$)}
                \State{$S_0$\texttt{.connect}($UE_{u}$, $eNB$)}
                \State{$u \gets u + 1$}
            \Until{$u = |U|$}
    \end{algorithmic}
    \end{footnotesize}
\end{algorithm}


\begin{algorithm}[!htb]
\caption{Variable Neighborhood Descent Local Search}
\label{alg:vndLocalSearch}
\begin{footnotesize}
    \begin{algorithmic}[1]
        \Require The user connected network $S_0$
        \Ensure The new optimized network $S$
            \State{$\mathcal{N} = [SwapUE, InsertUE]$}
            \State{$k \gets 1$} \Comment{Neighborhood structure selector}
            \State{$S \gets S_0$}
            \While{$k \leq |\mathcal{N}|$} 
                \State{$S' \gets \mathcal{N}_k(S)$} \label{alg:vnd:fi} \Comment{Find the first improvement neighbor $S'$ of $S$ in $\mathcal{N}_k$}
         	    \If{$Cost(S') < Cost(S)$}
         	        \State{$k \gets 1$} \label{alg:vnd:first1}
         	        \State{$S \gets S'$} \label{alg:vnd:first2}
         	     \Else
         	        \State{$k \gets k + 1$} \label{alg:vnd:next}
         	    \EndIf
            \EndWhile
            \State{\Return $S$}
    \end{algorithmic}
    \end{footnotesize}
\end{algorithm}

Algorithm~\ref{alg:vndLocalSearch} presents the pseudocode for the VND procedure, utilizing the first improvement strategy (Line \ref{alg:vnd:fi}), searching for a neighborhood of solution $S$ that improves the cost.
If the current neighborhood fails to improve the solution, it chooses a neighborhood from the set $\mathcal{N}$ systematically (line \ref{alg:vnd:next}) to continue the search. Upon improvement, the search progresses from the first neighborhood to the newly found solution (Lines \ref{alg:vnd:first1}--\ref{alg:vnd:first2}).

\section{Results}\label{sec:results}

In this section, we present our results for the ILP formulation~\eqref{equ:users}--\eqref{equ:rest8}, considering RSRQ and average handover frequency per eNB. We compare the results of our ILS-VND heuristic with the ILP formulation implemented using IBM ILOG CPLEX Optimization Studio V12.10~\cite{ibmcplex19}. CPLEX employs the branch and bound algorithm for optimal ILP solutions. We made the source code and data available in \cite{sourcecode2023}. This implementation does not consider a network simulator, but a similar scenario using the ns3 network simulator is provided in \cite{ahmadi:2020}. For evaluation instances, we followed the GAP standard generation scheme~\cite{Chu1997GeneticGAP} defined in Equations \eqref{equ:typeA}-\eqref{equ:typeB}. The number of UEs and eNBs is represented by $|U|$ and $|N|$, respectively. We kept the constant values such 0.6 and 0.7 from the generation scheme. Decreasing these values means that there is less bandwidth available in each eNB, which increases the difficulty for the models to find a solution, if there is one.
{\allowdisplaybreaks
    \begin{align}
        \textbf{Type A}: L_i = & 0.6\left( \frac{|U|}{|N|} \right)15+0.4L_{max} & \forall ~i \in N\label{equ:typeA}\\
        \textbf{Type B}: L_i = & 0.7(\text{\textbf{Type A}}) & \forall ~i \in N\label{equ:typeB}\\
        L_{max} = & \sum_{k \in U} {max(l_{ki}~|~i = 1, ...,n)} &\label{equ:rmax}
    \end{align}
}

\textbf{Type A} instances are easy, while \textbf{Type B} has tighter $L_i$ values for the eNBs bandwidth constraints. \textbf{Type B'} instances are similar to \textbf{Type B} but with more UEs (see Table~\ref{tbl:gapInstances}). We define user bandwidth requirements $l_{ik}$ as integers from a random uniform distribution U(5,25). We compute the maximum user bandwidth requirement for each eNB$_i$ and add them to obtain $L_{max}$, used to compute the bandwidth capacity $L_i$ of eNB$_i$.

\begin{table}[!htb]
    \centering
    \caption{The number of UEs and eNBs according to the instance type.}
    \label{tbl:gapInstances}
    \begin{tabular}{cccc} 
        \toprule
        \textbf{Instance} & \textbf{Type}        & \textbf{UEs Quantity} & \textbf{eNBs~\textbf{Quantity}}  \\ 
        \midrule
        1                 & \multirow{6}{*}{\textbf{A/B}} & 100                  & 5                               \\
        2                 &                      & 100                  & 10                              \\
        3                 &                      & 100                  & 20                              \\
        4                 &                      & 200                  & 5                               \\
        5                 &                      & 200                  & 10                              \\
        6                 &                      & 200                  & 20                              \\
        \midrule
        1                 & \multirow{6}{*}{\textbf{B'}}   & 1200                 & 5                               \\
        2                 &                      & 1200                 & 10                              \\
        3                 &                      & 1200                 & 20                              \\
        4                 &                      & 1500                 & 5                               \\
        5                 &                      & 1500                 & 10                              \\
        6                 &                      & 1500                 & 20                              \\
        \bottomrule
    \end{tabular}
\end{table}

\subsection{Heuristic Performance}

Regarding the heuristic, we execute the algorithms 30 times per instance on a computer with an Intel Core i7 @2.7 GHz x 4 processor, 16 GB RAM, and Ubuntu 18.04 LTS. We use $T = 2$ seconds and $\alpha = 2$ in Algorithm~\ref{alg:ils}. The average execution time for the ILP formulation and the ILS-VND heuristic are 61.43 ms and 11.09 ms, respectively, showing an 82\% reduction. Table~\ref{tbl:heuristicResults2} displays the results, containing solution value and the average execution time. The ILS-VND found the optimal solution provided by the ILP model for all instances, in each of the 30 executions. However, heuristic algorithms may not always provide the optimal solution, and its use should be analyzed according to each problem \cite{desale2015heuristic}. The major advantage of our heuristic algorithm is its ability to find answers much faster, which makes it suitable for online deployment in a mobile network.


\begin{center}

\input{tables/allAlgResult2}
\end{center}

\subsection{Mobility Simulation Scenario}

\begin{figure}[!htb]
    \centering
    \includegraphics[width=.9\linewidth]{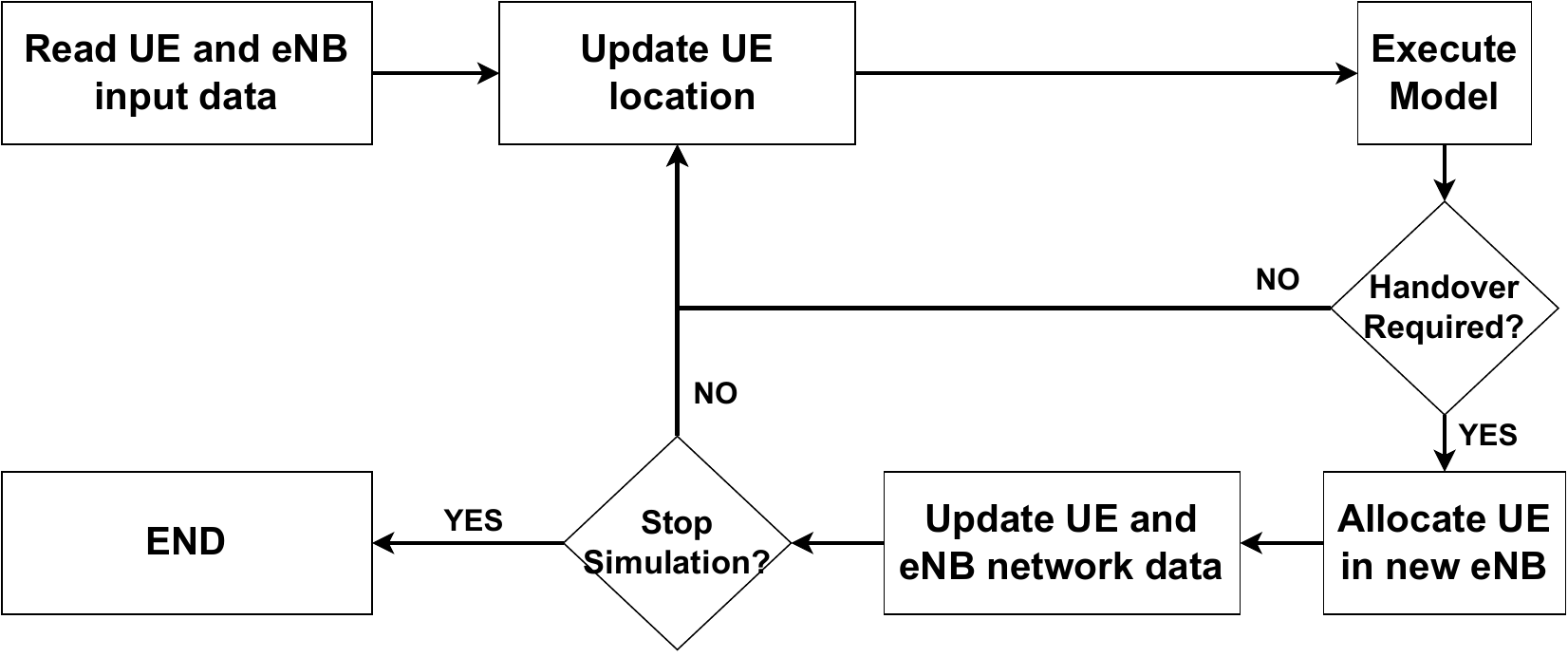}
    \caption{Simulation flowchart.}
    \label{fig:flowchart}
\end{figure}

To obtain the results in this section, we execute the ILS-VND metaheuristic in a vehicular mobility simulation scenario. We design the model following the flowchart depicted in Fig.~\ref{fig:flowchart}. At the initial step, the UE can be connected to any base station, and the final network solution is not limiting eNB communication range. This will be addressed in future work. Input data includes eNBs handover frequency, locations, UE and eNB bandwidth requirements. UE location updates and association follows Eqs. \eqref{equ:users}--\eqref{equ:rest8}.
Finally, the model performs the handover if a better eNB, in terms of our objective function, is available to a given UE. Otherwise, the algorithm collects and updates information about eNBs and UEs and runs periodically.
Furthermore, we must connect every user to one base station while keeping network service and bandwidth limits. 
We solve this association problem using the ILP formulation and the proposed heuristic algorithm.

\begin{figure}[!htb]
    \centering
    \includegraphics[width=.7\linewidth]{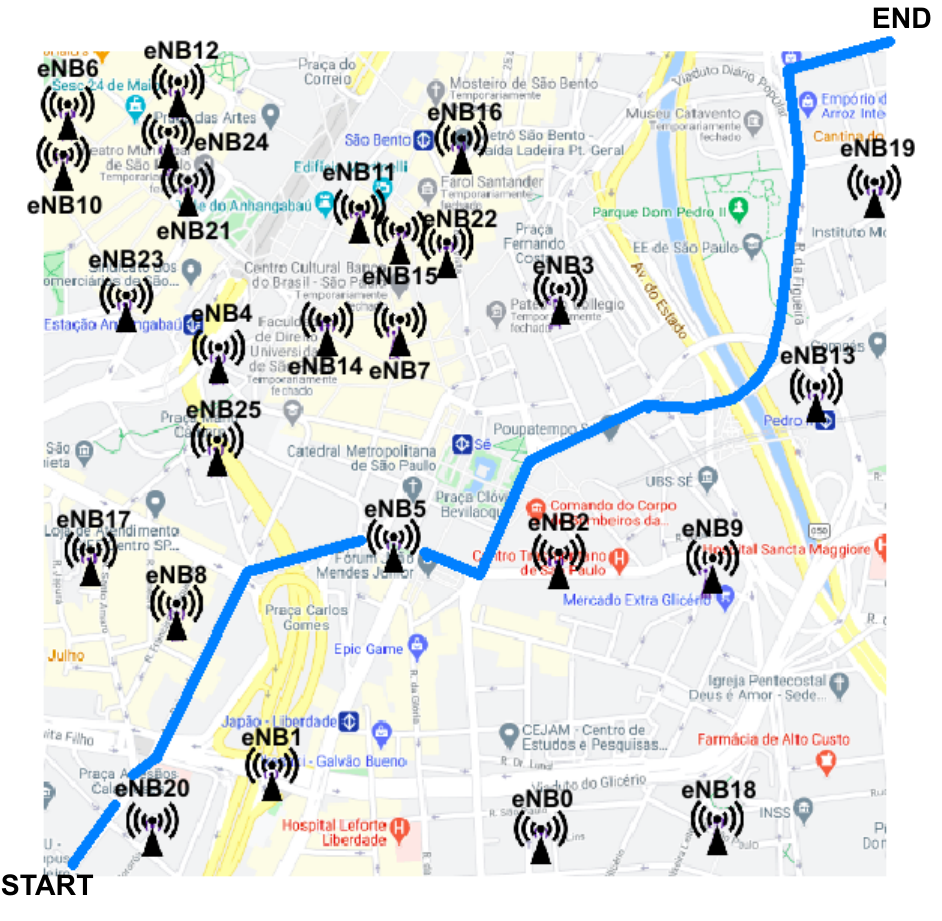}
    \caption{Simulated vehicular UE route in São Paulo city, Brazil, considering real-world eNB locations.}
    \label{fig:spRouteSimulation}
\end{figure}

This experiment simulates the vehicular UE association considering eNBs in São Paulo, Brazil (Fig.~\ref{fig:spRouteSimulation}). Real-world eNB locations are provided by Telebrasil~\cite{telebrasil:2022}. The region of interest is $(1947.65 \times 1878.95)$ m$^2$ with $26$ eNBs from a local carrier. Each eNB, labeled from $0$ to $25$, has a transmission range of $R_{i} = 500$ m. Using the SUMO simulator~\cite{Krajzewicz2012SUMO}, we generate a route (highlighted in blue in Fig.~\ref{fig:spRouteSimulation}) that includes $432$ UE location readings.

We also compare our ILS-VND user association model with the solution proposed by Ahmadi et al.~\cite{ahmadi:2020}, through an illustrative example detailed in the follwoing. In~\cite{ahmadi:2020}, the authors construct an eNB score rank based on route prediction, assigning a score to each eNB according to:
\begin{align}
    score &= RSRQ + Factor_{predict} 
    \label{equ:ahmadi}
\end{align}
where the prediction factor is defined by:
\begin{align}
    Factor_{predict} &= factor_{direction} +factor_{distratio} \label{equ:ahmadi2}
\end{align}

In their solution, if a user approaches a specific eNB and improves the RSRQ value simultaneously, then $factor_{direction} = 1$, and $0$ otherwise. The $factor_{distratio}$ depends on the user current location and all potential future routes. The eNB that covers the most probable future routes receives the highest score. As for the RSRQ parameter, a more robust signal leads to a higher score. 

\begin{figure}[!htb]
    \centering
    \includegraphics[width=.7\linewidth]{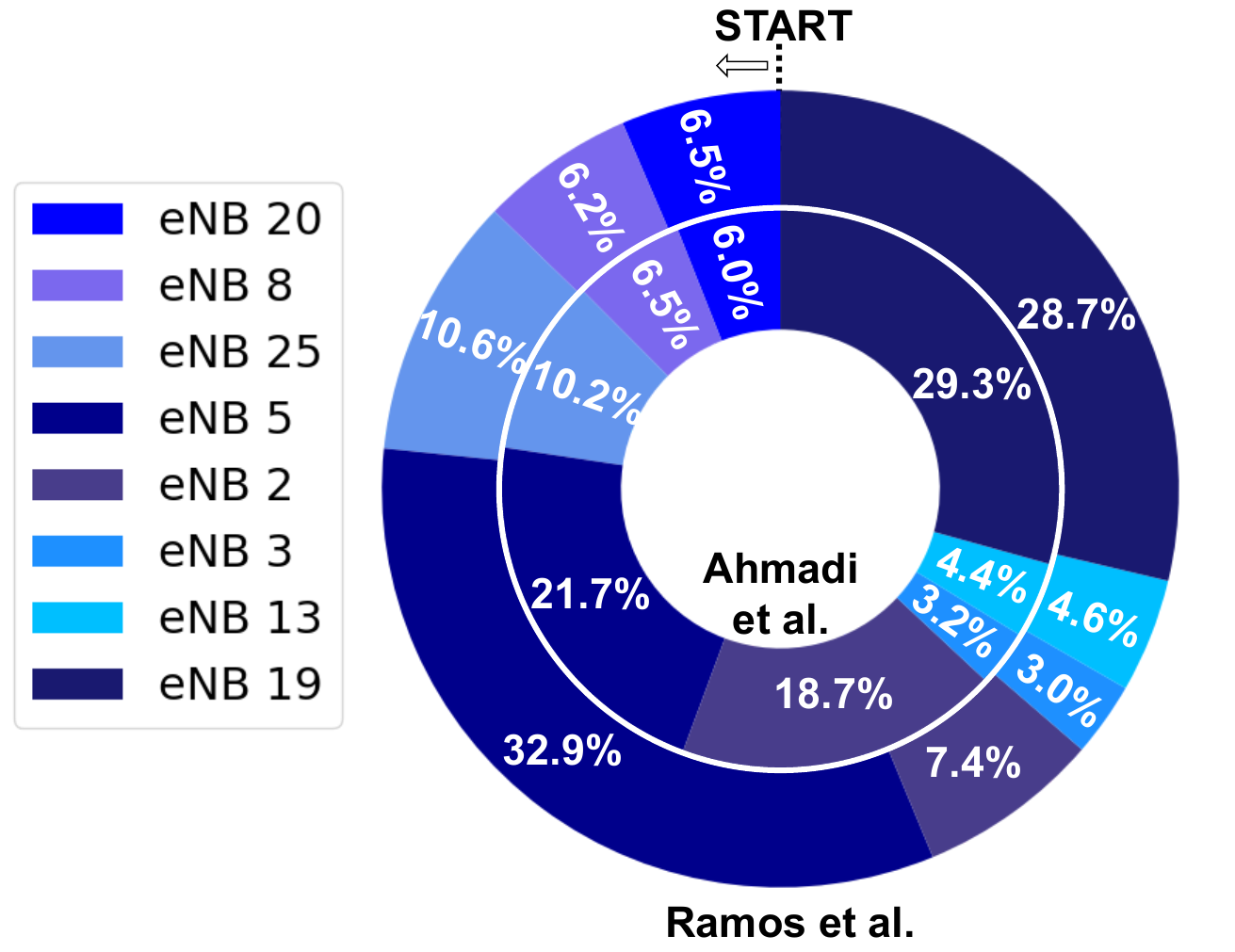}
    \caption{The allocation results for the ILS-VND (outer) and Ahmadi et al. (inner) models.}
    \label{fig:alloResults}
\end{figure}

Fig.~\ref{fig:alloResults} illustrates the user association results for each of these approaches. The mobility simulation process reads one location per iteration until the final destination. The different models receive the locations to compute the best candidate eNB. For each solution, the UE starts at eNB$_{20}$ and executes seven handovers, following the user association sequence 20$\rightarrow$8$\rightarrow$25$\rightarrow$5$\rightarrow$2$\rightarrow$3$\rightarrow$13$\rightarrow$19. Although both models perform the same associations, user connection duration vary due to RSRQ value influence, affecting both Ahmadi et al.~\cite{ahmadi:2020} and our solution.

\begin{figure}[!htb]
    \centering
    \includegraphics[width=\linewidth]{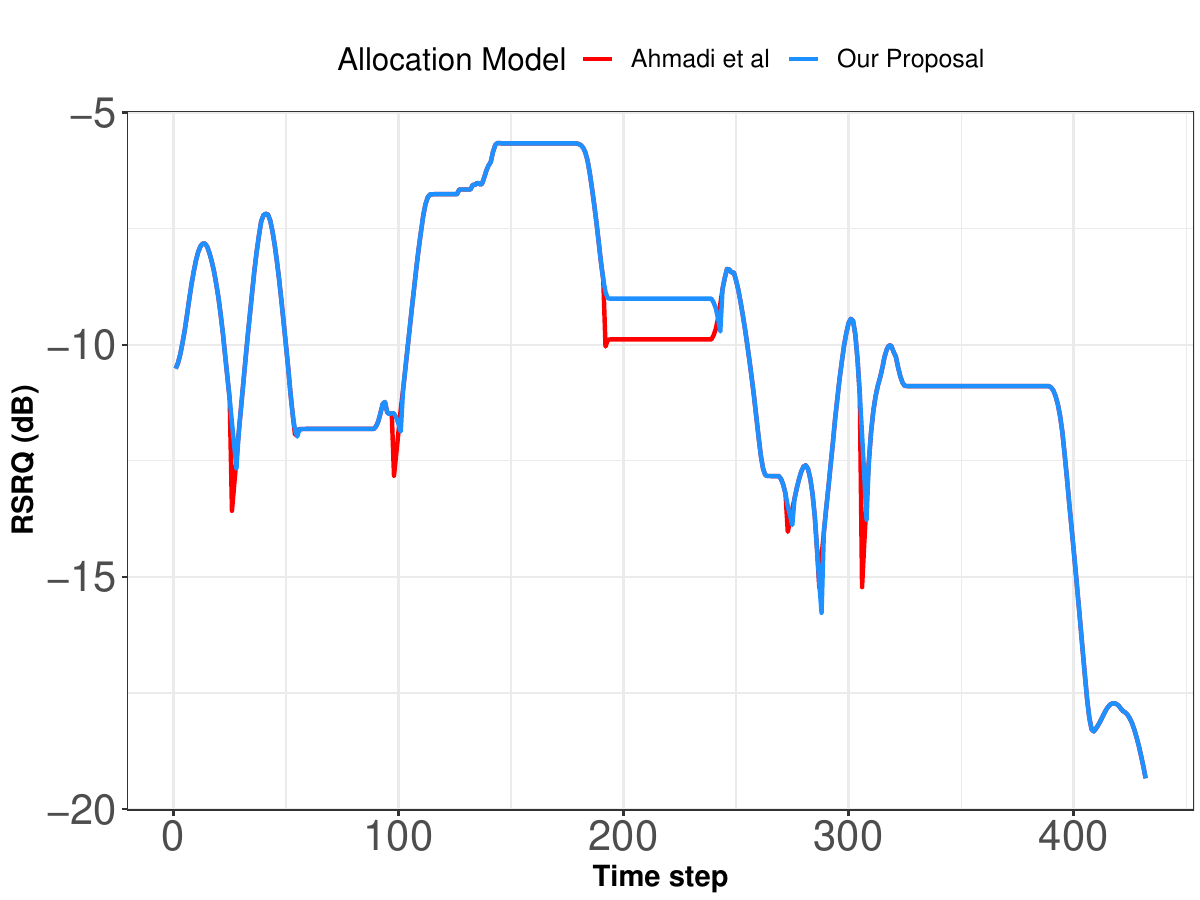}
    \caption{User link RSRQ value through the route.}
    \label{fig:heuRSSQPlot}
\end{figure}

Fig.~\ref{fig:heuRSSQPlot} shows the user RSRQ status. Optimal RSRQ eNB selection is crucial to maintain connection quality, with values closer to zero being better. Results mostly overlap, but our model outperforms occasionally. For example, in our model, UE remained connected to eNB$_5$ for 32.9\% of the route, compared to 21.9\% using~\cite{ahmadi:2020}. This result is visible around the $200$ time-step mark in Fig.~\ref{fig:heuRSSQPlot}. In another example, our heuristic maintained the UE connection to eNB$_5$ for 7.4\% of the route, while~\cite{ahmadi:2020} achieved 18.7\%. Ideally, user association models should prioritize longer user connectivity, while maintaining service quality and reducing handovers. If we consider the number of times each model successfully maintained user connectivity for longer at each eNB compared to the other, both models achieved this four times.

The critical difference between the two approaches is that we only consider the average handover frequency $h_i$ and the RSRQ value $\Theta_{ki}$ to compute the best candidate eNB. At the same time,~\cite{ahmadi:2020} is doing route prediction, which requires data from all the possible routes connecting a user's current location and final destination. We implemented the simulation proposed in~\cite{ahmadi:2020} to consider the single-path-best-case scenario of Fig.~\ref{fig:spRouteSimulation}, since multiple routes can influence the ideal point choice and negatively affect their solution. Our solution was still able to achieve similar results, even without including any prediction function. If human mobility is indeed highly predictable, as claimed by some studies~\cite{song2010limits}, our heuristic can incorporate route prediction features and reduce the number of handovers. For example, in our toy scenario, if we have data of previous association results and know that the UE is likely to follow the route in Fig.~\ref{fig:spRouteSimulation} weekly, at the same hour, we can pre-configure offline the eNB sequence 20$\rightarrow$8$\rightarrow$5$\rightarrow$2$\rightarrow$13$\rightarrow$19, while keeping the UE QoS level. This strategy would remove two handover operations.

\begin{figure}[!htb]
    \centering
    \includegraphics[width=\linewidth]{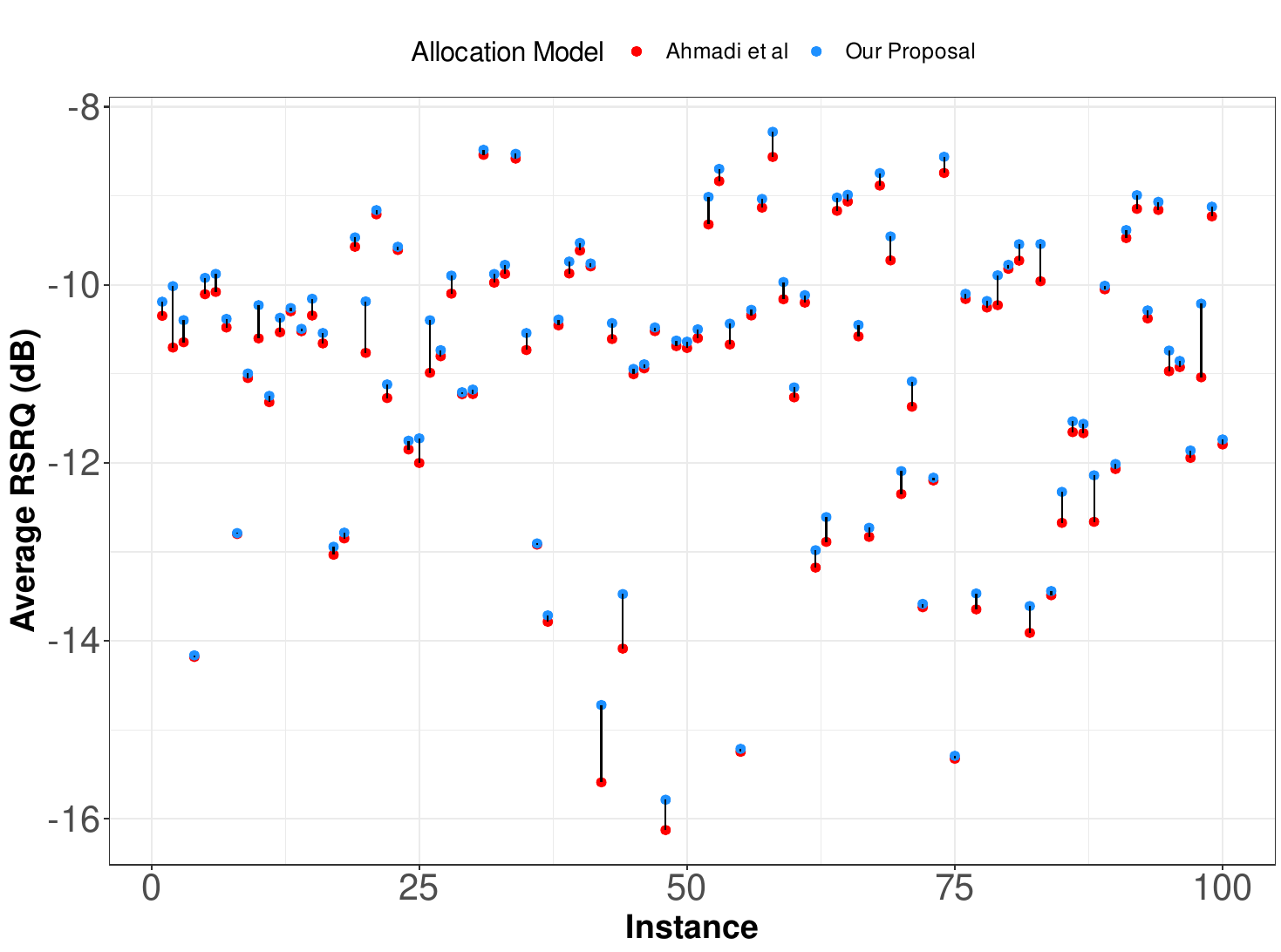}
    \caption{The average RSRQ value for each instance's route.}
    \label{fig:routeAvgRSRQ}
\end{figure}

To investigate if the previous result is not solely an artifact of base station placement, we generated 100 instances with different random eNB locations. We maintained 26 eNBs, along with the area and UE route. With the RSRQ time series for each individual run, we calculate the average value, shown in Fig.~\ref{fig:routeAvgRSRQ}. The overall RSRQ averages for Ahmadi et al.~\cite{ahmadi:2020} and our proposal are -11.03 dB and -10.87 dB, respectively, resulting in a 1.45\% gain for our association model. A two-sided Kolmogorov-Smirnov test with a 5\% significance level provides a $p$-value of 0.3667 ($>$ 0.05), reinforcing that the models are statistically equivalent. 

However, as explained, the multi-route model presented by~\cite{ahmadi:2020} relies on third-party applications to compute all possible routes for every network user to reach the final destination. Finding all possible paths between two points in this way, for each user, is expensive. Besides, this approach may face significant response time overhead due to third-party applications network delays, being more suited for less dynamic scenarios, such as public transportation routes. Therefore, we see our proposal as more flexible and suitable for general transportation systems.

\section{Conclusion}\label{sec:conclusion}

This paper presents the ILS-VND heuristic for vehicular user association and handover decision in mobile networks. We achieved optimal solutions with an 82\% reduction in execution time compared to a linear programming formulation. When comparing our solution to a concurrent user association model that considers route prediction, we achieved an average gain of 1.45\% in terms of signal quality, while maintaining the handover count and requiring a much lower level of input information. Future work includes studying route prediction, and exploring new optimization parameters. Additionally, we can explore a hybrid approach by combining ILS-VND with an exact method by generating initial solutions using heuristics and utilizing them as input for the ILP formulation.

\section*{Acknowledgments}

We thank the Research Foundation of the State of Alagoas (FAPEAL) under grants E:60030.0000000352/2021 and the National Council for Scientific and Technological Development (CNPq) under grant 407515/2022-4.

\bibliographystyle{IEEEtran}
\bibliography{main}

\vfill

\end{document}

%% file: tables/qualitative_table.tex
\begin{table}[!htb]
	\centering
	\caption{Comparison between main features of our work and similar literature proposals.}
	\label{tbl:qualitative_analysis}
	\resizebox{\linewidth}{!}{
		\begin{tabular}{@{}cccccc@{}}
			\toprule
			Authors                                             & QoS & BM & MS & UAM & HS \\ \midrule
			Proposed solution                                   & \checkmark         & \checkmark           & \checkmark          & \checkmark            & \checkmark      \\
			Ahmadi et al. (2020) \cite{ahmadi:2020}             & \checkmark         & --                    & \checkmark          & \checkmark            & --               \\
			Bakht et al. (2019) \cite{bakht2019powerAllocation} & \checkmark         & \checkmark           & --                   & \checkmark            & --               \\
			Lee et al. (2017) \cite{jiseong:2017}               & \checkmark         & \checkmark           & \checkmark          & \checkmark            & --               \\
			Taleb et al. (2015) \cite{tarik2015}                & \checkmark         & \checkmark           & \checkmark          & --                     & --               \\ \bottomrule
		\end{tabular}
	}
\end{table}

%% file: tables/allAlgResult2.tex
\begin{table}[!htb]
\centering
\caption[Exact and heuristic models result.]{The results for the ILS, and the exact model.}
\label{tbl:heuristicResults2}
\begin{tabular}{cccccc}
\hline
\multicolumn{1}{l}{\multirow{2}{*}{\textbf{Inst.}}} & \multirow{2}{*}{\textbf{Type}}        & \multicolumn{2}{c}{\textbf{ILP-Formulation}} & \multicolumn{2}{c}{\textbf{ILS-VND}} \\ \cline{3-6} 
\multicolumn{1}{l}{}                       &                              & \textbf{Best}            & \textbf{Time (ms)}         & \textbf{Best}        & \textbf{Time (ms)}     \\ \hline
1                                          & \multirow{6}{*}{\textbf{A}}  & 101.77          & 5.80              & 101.77      & 0.18          \\
2                                          &                              & 94.82           & 9.65              & 94.82       & 0.21          \\
3                                          &                              & 89.17           & 16.34             & 89.17       & 0.24          \\
4                                          &                              & 209.67          & 8.36              & 209.67      & 0.72          \\
5                                          &                              & 179.97          & 17.46             & 179.97      & 0.67          \\
6                                          &                              & 169.15          & 31.09             & 169.15      & 1.02          \\ \hline
1                                          & \multirow{6}{*}{\textbf{B}}  & 102.64          & 5.35              & 102.64      & 0.19          \\
2                                          &                              & 91.96           & 8.68              & 91.96       & 0.22          \\
3                                          &                              & 88.38           & 19.37             & 88.38       & 0.26          \\
4                                          &                              & 235.75          & 7.91              & 235.75      & 0.88          \\
5                                          &                              & 197.79          & 19.53             & 197.79      & 0.77          \\
6                                          &                              & 158.65          & 34.29             & 158.65      & 1.00          \\ \hline
1                                          & \multirow{6}{*}{\textbf{B'}} & 1221.85         & 47.87             & 1221.85     & 23.57         \\
2                                          &                              & 1057.96         & 109.02            & 1057.96     & 25.15         \\
3                                          &                              & 1020.05         & 237.66            & 1020.05     & 26.88         \\
4                                          &                              & 1636.13         & 76.29             & 1636.13     & 37.10         \\
5                                          &                              & 1397.80         & 143.22            & 1397.80     & 38.28         \\
6                                          &                              & 1248.36         & 307.77            & 1248.36     & 42.25         \\ \hline
\end{tabular}
\end{table}

%% file: main.bbl
\begin{thebibliography}{10}
\providecommand{\url}[1]{#1}
\csname url@samestyle\endcsname
\providecommand{\newblock}{\relax}
\providecommand{\bibinfo}[2]{#2}
\providecommand{\BIBentrySTDinterwordspacing}{\spaceskip=0pt\relax}
\providecommand{\BIBentryALTinterwordstretchfactor}{4}
\providecommand{\BIBentryALTinterwordspacing}{\spaceskip=\fontdimen2\font plus
\BIBentryALTinterwordstretchfactor\fontdimen3\font minus \fontdimen4\font\relax}
\providecommand{\BIBforeignlanguage}[2]{{%
\expandafter\ifx\csname l@#1\endcsname\relax
\typeout{** WARNING: IEEEtran.bst: No hyphenation pattern has been}%
\typeout{** loaded for the language `#1'. Using the pattern for}%
\typeout{** the default language instead.}%
\else
\language=\csname l@#1\endcsname
\fi
#2}}
\providecommand{\BIBdecl}{\relax}
\BIBdecl

\bibitem{ramos20195Gsdn}
G.~S. Ramos, R.~G.~S. Pinheiro, and A.~L.~L. Aquino, ``Optimizing 5g networks processes with software defined networks,'' in \emph{2019 IEEE 8th International Conference on Cloud Networking (CloudNet)}.\hskip 1em plus 0.5em minus 0.4em\relax IEEE, 2019, pp. 1--6.

\bibitem{tarik2015}
T.~Taleb, M.~Bagaa, and A.~Ksentini, ``{User mobility-aware Virtual Network Function placement for Virtual 5G Network Infrastructure},'' in \emph{2015 IEEE International Conference on Communications (ICC)}, 2015.

\bibitem{ahmadi:2020}
K.~Ahmadi, S.~P. Miralavy, and M.~Ghassemian, ``Software-defined networking to improve handover in mobile edge networks,'' \emph{International Journal of Communication Systems}, p. e4510, jun 2020.

\bibitem{sourcecode2023}
\BIBentryALTinterwordspacing
G.~S. Ramos, R.~Stanica, R.~G.~S. Pinheiro, and A.~L.~L. Aquino. (2023, oct) Users association source code. Accessed on october 2023. [Online]. Available: \url{https://github.com/Geymerson/mobile_user_association}
\BIBentrySTDinterwordspacing

\bibitem{xu2021survey}
Y.~Xu, G.~Gui, H.~Gacanin, and F.~Adachi, ``A survey on resource allocation for 5g heterogeneous networks: Current research, future trends, and challenges,'' \emph{IEEE Communications Surveys \& Tutorials}, vol.~23, no.~2, pp. 668--695, 2021.

\bibitem{jiseong:2017}
J.~Lee and Y.~Yoo, ``{Handover Cell Selection Using User Mobility Information in a 5G SDN-based Network},'' \emph{{2017 Ninth International Conference on Ubiquitous and Future Networks (ICUFN)}}, pp. 697--702, jul 2017.

\bibitem{bakht2019powerAllocation}
K.~Bakht, F.~Jameel, Z.~Ali, W.~U. Khan, I.~Khan, G.~A. Sardar~Sidhu, and J.~W. Lee, ``Power allocation and user assignment scheme for beyond 5g heterogeneous networks,'' \emph{Wireless Communications and Mobile Computing}, vol. 2019, 2019.

\bibitem{caroe1999dualDecomp}
C.~C. Car{\o}e and R.~Schultz, ``Dual decomposition in stochastic integer programming,'' \emph{Operations Research Letters}, vol.~24, no. 1-2, pp. 37--45, 1999.

\bibitem{Chatterjee2014PossiblePaths}
S.~Chatterjee and D.~Banerjee, ``{A Novel Boolean Expression based Algorithm to find all possible Simple Paths between two nodes of a Graph.}'' \emph{International Journal of Advanced Research in Computer Science}, vol.~5, no.~7, 2014.

\bibitem{cattrysse1992survey}
D.~G. Cattrysse and L.~N. Van~Wassenhove, ``A survey of algorithms for the generalized assignment problem,'' \emph{European journal of operational research}, vol.~60, no.~3, pp. 260--272, 1992.

\bibitem{Lourencco2019Iterated}
H.~R. Louren{\c{c}}o, O.~C. Martin, and T.~St{\"u}tzle, ``Iterated local search: Framework and applications,'' in \emph{Handbook of metaheuristics}.\hskip 1em plus 0.5em minus 0.4em\relax Springer, 2019, pp. 129--168.

\bibitem{Hansen_2018}
P.~Hansen, N.~Mladenovi{\'{c}}, J.~Brimberg, and J.~A.~M. P{\'{e}}rez, ``Variable neighborhood search,'' in \emph{Handbook of Metaheuristics}.\hskip 1em plus 0.5em minus 0.4em\relax Springer International Publishing, sep 2018, pp. 57--97.

\bibitem{matsumoto1998mersennetwister}
M.~Matsumoto and T.~Nishimura, ``{Mersenne Twister: A 623-Dimensionally Equidistributed Uniform Pseudo-random Number Generator},'' \emph{ACM Transactions on Modeling and Computer Simulation (TOMACS)}, vol.~8, no.~1, pp. 3--30, 1998.

\bibitem{ibmcplex19}
IBM, ``{IBM ILOG CPLEX Optimization Studio Getting Started with CPLEX},'' \url{https://www.ibm.com/docs/en/SSSA5P_12.8.0/ilog.odms.studio.help/pdf/gscplex.pdf}, apr 2017, online, retrieved May 01, 2022.

\bibitem{Chu1997GeneticGAP}
P.~C. Chu and J.~E. Beasley, ``{A Genetic Algorithm for the Generalised Assignment Problem},'' \emph{Computers \& Operations Research}, vol.~24, no.~1, pp. 17--23, 1997.

\bibitem{desale2015heuristic}
S.~Desale, A.~Rasool, S.~Andhale, and P.~Rane, ``Heuristic and meta-heuristic algorithms and their relevance to the real world: a survey,'' \emph{Int. J. Comput. Eng. Res. Trends}, vol. 351, no.~5, pp. 2349--7084, 2015.

\bibitem{telebrasil:2022}
Telebrasil, ``{Mapa de ERBs Brasil (antenas)},'' \url{http://www.telecocare.com.br/telebrasil/erbs/}, online, retrieved May 01, 2022.

\bibitem{Krajzewicz2012SUMO}
D.~Krajzewicz, J.~Erdmann, M.~Behrisch, and L.~Bieker, ``{Recent Development and Applications of SUMO-Simulation of Urban MObility},'' \emph{International journal on advances in systems and measurements}, vol.~5, no. 3\&4, 2012.

\bibitem{song2010limits}
C.~Song, Z.~Qu, N.~Blumm, and A.-L. Barab{\'a}si, ``Limits of predictability in human mobility,'' \emph{Science}, vol. 327, no. 5968, pp. 1018--1021, 2010.

\end{thebibliography}
